\documentclass[referee]{cjaa}           

\usepackage{graphicx}                   
\input{epsf.sty}                        
\input{psfig.sty}                       

\begin{document}

   \title{Class Transitions in Black Holes}


   \volnopage{Vol.0 (200x) No.0, 000--000}      
   \setcounter{page}{1}          

   \author{Sandip K. Chakrabarti
      \inst{1,2}\mailto{}
      }

   \institute{S. N. Bose National Centre for Basic Sciences, Salt Lake, Kolkata, 700098, India\\
             \email{chakraba@bose.res.in}
\and
Centre for Space Physics, Chalantika 43, Garia Station Road, Kolkata 700084, India\\ }

   \date{Received~~2004 Sept. 10; accepted~~2004~~month day}

\abstract{A black hole spectrum is known to change from the hard state to  the soft state when the energy
spectral index $\alpha$ ($F_E \propto E^{-\alpha}$) in, say, $2-20$ keV range changes
from $\alpha \sim 0.5$ to $\sim 1.5$. However, this `classical' definition which characterizes 
black holes like Cyg X-1, becomes less useful for many objects such as GRS 1915+105 
in which the spectral slope is seen to vary from one to the other in a matter of seconds
and depending on whether or not winds form, the spectral slope also changes.
The light curves and the colour-colour diagrams may look completely different 
on different days depending on the frequency and mode of switching from one 
spectral state to the other. Though RXTE observations have yielded wealth of 
information on such `variability classes' in GRS 1915+105, very rarely one 
has been able to observe how the object goes from one class to the other. 
In the present review, we discuss possible origins of the class transition 
and present several examples of such transitions. In this context, we use 
mostly the results of the Indian X-ray Astronomy Experiment (IXAE) which 
observed GRS 1915+105 more regularly.
}
   \authorrunning{ S. K. Chakrabarti}            
   \titlerunning{Class Transitions of Black holes}  

   \maketitle

\noindent To be Published in the proceedings of the 5th Microquasar Conference: 
Chinese Journal of Astronomy and Astrophysics


%
%
\section{Introduction}           
\label{sect:intro}

Black holes are fundamentally `black' in that they cannot be observed directly.
However, the behaviour of radiation emitted by the matter falling onto them 
can be studied through carefully planned observations and a wealth of 
information about the black hole and the accretion flow could be obtained.
In the present review, we concentrate on the characteristics of class transitions 
in black holes and infer possible physical processes which may be responsible for
such transitions. 

\section{Nature of emitted radiation: theoretical expectations}

Radiations emitted from the infalling matter carry information about the 
density, temperature and magnetic field distribution. Theoretically, these flow variables are derived 
in conjunction with the velocity profiles (subsonic/supersonic nature) when proper hydro- and magnetohydro- dynamic
treatments are made.  By `proper', we mean by actually solving full set of equations 
with appropriate boundary conditions and not making `models'. The general picture which comes out has been discussed in several
papers and we refer to the reviews Chakrabarti (1996, 2001) in this connection.

In order to re-iterate the importance of the role of transonic solutions in understanding 
black hole accretion process and therefore the spectral variations including the 
variability class transitions, we present in Fig. 1(a-b) two types of basic solutions of the viscous
transonic flows (see, Chakrabarti, 1990, 1996 for details). In Fig. 1a, (upper panel) a solution with a small viscosity
parameter is presented which includes a standing shock transition shown by a vertical arrow. In the left, 
the Mach number variation is presented and in the right, the flow configuration is presented. 
In the outer regions, when the flow is sub-sonic, it is more or less Keplerian (K). After the flow 
becomes super-sonic, the flow becomes sub-Keplerian (SK). It passes through a centrifugal force
induced standing or oscillating shock and become subsonic again which enters through the inner sonic point
to become super-sonic. This region between the shock and the horizon is known as the  
CENtrifugal pressure dominated BOundary Layer or CENBOL which essentially releases most of the observable 
X-rays in a galactic black hole accretion. When the viscosity is higher (Fig. 1b), the solution changes its
character: there are now {\it two} solutions, one directly coming from a Keplerian disk and entering through the
inner sonic point (lower branch), while the other passes through the outer sonic point before disappearing into a black hole.
The configuration profiles show the essential differences. It is natural that both of these steady solutions
cannot be realized simultaneously. Time dependent simulations always prefer the higher entropy  
solution passing through the inner sonic point. Thus while at lower viscosities, the flow
forms a steady or oscillating shock, at higher viscosities the flow indeed forms a shock by staying on the
upper branch, but the shock propagates 
away to a larger distance (Chakrabarti \& Molteni, 1995) while converting the post-shock region 
into a Keplerian disk on its way. Thus, it ends up in the lower branch. 
This is how a Keplerian disk forms in the first place. Fig. 2(a-b) shows this behaviour where evolution of the Mach 
number and the specific angular momentum are shown. 

\begin{figure}
   \begin{center}
{\vskip 0.0cm
   \mbox{\epsfxsize=0.8\textwidth\epsfysize=0.6\textwidth\epsfbox{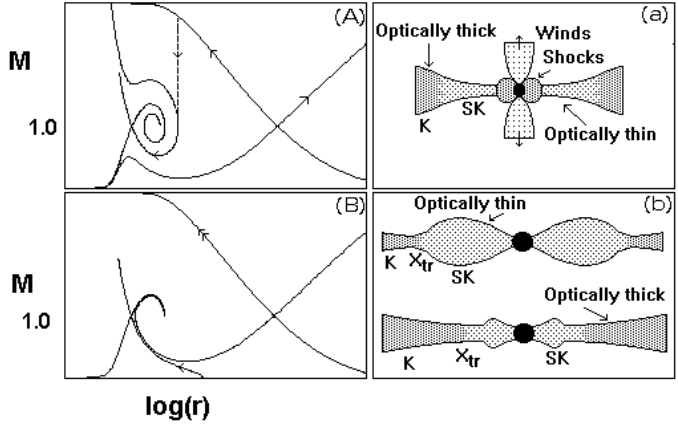}}}
\vskip -7.0cm
   \caption{Typical solutions of viscous transonic flows are shown in the left panels
for (a) viscosity parameter lower than the critical value and (b) viscosity parameter above
the critical value. The corresponding flow configurations are shown on the right.
}
   \end{center}
\end{figure}

\begin{figure}
   \begin{center}
{\vskip 0.0cm
   \mbox{\epsfxsize=0.7\textwidth\epsfysize=0.5\textwidth\epsfbox{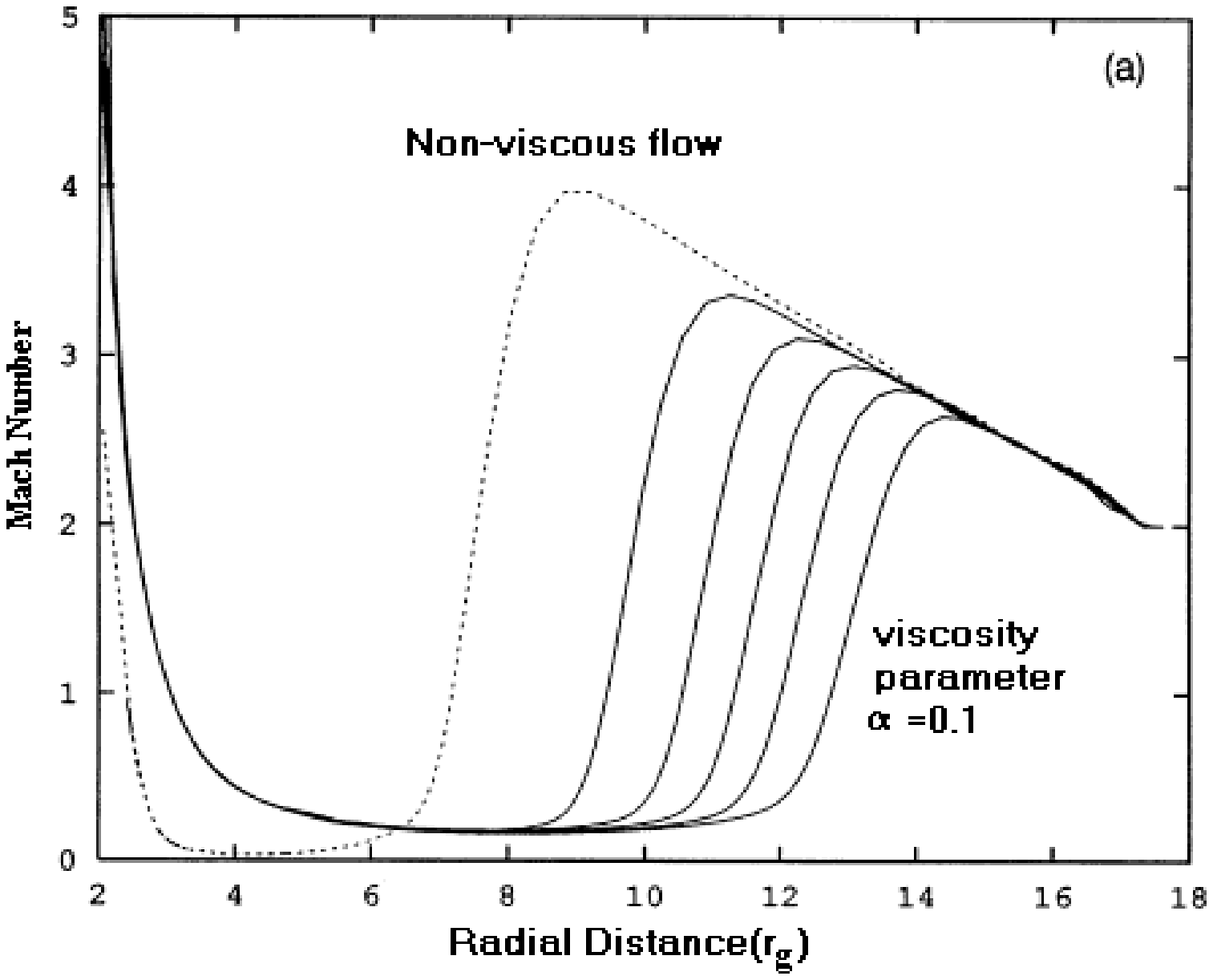}}
\vskip 0.0 cm
\hskip -1.0cm
\mbox{\epsfxsize=0.75\textwidth\epsfysize=0.55\textwidth\epsfbox{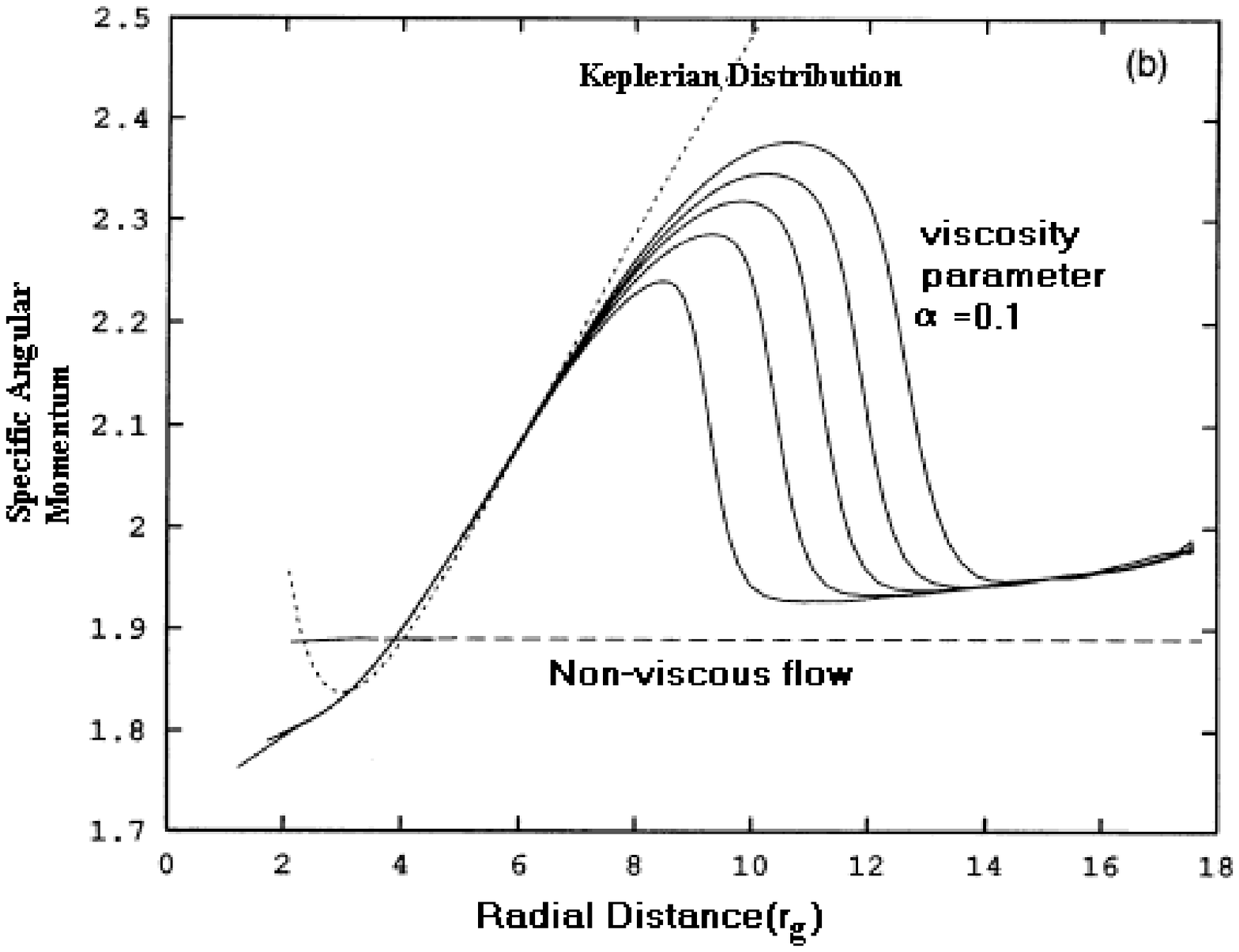}}
}
\vskip 0.0cm
   \caption{Time dependent solutions of (a) the Mach number and (b) the specific angular momentum distribution 
of an advective flow in which viscosity parameter is above the critical value. The shock, formed closed to the
black hole, propagates outward leaving behind a Keplerian disk between the shock and marginally stable
orbit. Solutions without viscosity are also placed for comparison.}
   \end{center}
\end{figure}

Thus we observe that both the Keplerian and the sub-Keplerian flow could be present depending on viscosity parameter
and on the basis of these theoretical considerations, Chakrabarti (1994) and Chakrabarti \& Titarchuk (1995) 
proposed that perhaps both the components are present in a generic flow around galactic and extragalactic black
holes with the Keplerian disk in the equatorial plane flanked by sub-Keplerian flows above and below. 
Subsequently, it was found that the CENBOL could be the source of the jets as well (Chakrabarti, 
1998, 1999) and the ratio of the outflow rate to the inflow rate $R_{jet}={\dot m}_{out}/{\dot m}_{in}$ strongly 
depends on the strength of the shock, characterized by the compression ratio $R$. In both the weak 
($R\sim 0$) and the strong ($R\sim 4-7$) shocks, the ratio is low, while for the intermediate 
shocks of strength ($\sim 2-3$) the ratio is very high. A high $R_{jet}$, coupled to a high ${\dot m}_{in}$ 
produces profuse amount of outflows which can influence the spectral properties in the following way:
(a) the outflow, up to the sonic sphere, moves slowly and has higher optical depth. It intercepts the 
soft-photons from the disk and reprocesses them. If the optical depth is high enough, it can be cooled efficiently
and the flow might just loss its drive to go out and fall back into the disk itself, temporary increasing the 
accretion rates. Eventually, after this temporarily phase is over after the viscous time scale of a few 
tens of seconds, the CENBOL develops once more and the outflow starts all over again.
All these phases could be observable in the spectral feature and light curves which
determine the `class' of a black hole. (b) In case the accretion rate is not high enough, the outflow
will not be cooled and would continue to propagate steadily. This type of `class' may continue for a
long time, unless the rates (Keplerian and sub-Keplerian) themselves are fundamentally changed at the outer edge
in the first place. Change in viscosity brings rise to changes in the Keplerian/sub-Keplerian
rates (see Fig. 2 above) as well. 

According to this Two Component Advective Flow (TCAF) paradigm, oscillations of shocks give rise 
to the quasi-periodic oscillations (QPOs). Thus QPOs are possible when shocks are. But the 
reverse is not true, since not all the shocks are expected to oscillate (Molteni, Sponholz \& Chakrabarti, 1996;
Chakrabarti, Acharyya \& Molteni, 2004). Shocks are also useful in accelerating particles and producing 
non-thermal spectra. Thus, high energy emissions may also be explained from by TCAF solution (Chakrabarti, 2004)
The existence of two components in the flow has been verified observationally (Smith, Heindl \& Swank, 2002). 

In Fig. 3, we present a few types of configurations of the Two Component Advective Flows (TCAF)
which give rise to different types of light curves. In the upper left, the sub-Keplerian component has no sharp
shock transition, but the flow is puffed up due to slowing down of matter (and becoming hotter
in the process) due to the  centrifugal barrier. This is expected to 
produce a light curve with no QPO and with a low rate of outflow. The spectrum will be harder. In upper-right,
we schematically show the CENBOL as a spherical region (which in reality is a `thick disk' like region
with funnel etc. (see, Chakrabarti, 1993) which is formed due to a steady or an oscillating shock. 
Thermally/centrifugally/magneto-centrifugally driven wind comes out of this region.  This configuration is
expected to produce harder spectrum with a QPO with possible drifting frequency due to variation in accretion rate 
and viscosity. Accretion rate controls the cooling rate and hence the shock (QPO) oscillation frequency.
In the lower left, we show the situation when the outflow could become optically thick and return flow 
causes the jets to be blobby. Return flow temporarily increases the accretion rate and soft-state
may develop. This may cause switching of states at the interval of a few seconds in some of the classes.
In lower-right, the viscosity is so high that the Keplerian disk reaches close to the inner sonic point (Fig. 2b)
and in this soft/quasi-soft state only hard spectra is produced by the bulk-motion Comptonization and non-thermal
particles in accretion and wind shocks.

It is to be noted that in the TCAF solution, the advective flow itself is the Compton cloud.
The literature is overwhelmed with the models such as ADAF, dynamical corona and the like 
which can at the best mimic the CENBOL of TCAF solution. A standing, propagating or oscillating 
shock close to a black hole, is not a wishful thinking, but it is a reality. The cartoon diagram may remind 
people of two-temperature flow of Shapiro, Lightman and Eardley (1976), but the difference is that presently
one is not discussing `Keplerian' advective flow as no such thing exists. Similarly, it may also remind
one of thick accretion flow: this is also not correct, as the thick disk solutions are purely rotating
and are not attached to any disk. The only self-consistent solutions which include thick disk like 
feature as well as the disk is the advective flow model presented more than ten years back (see, Chakrabarti, 1993).

\begin{figure}
   \begin{center}
{\vskip 0.0cm
   \mbox{\epsfxsize=0.8\textwidth\epsfysize=0.8\textwidth\epsfbox{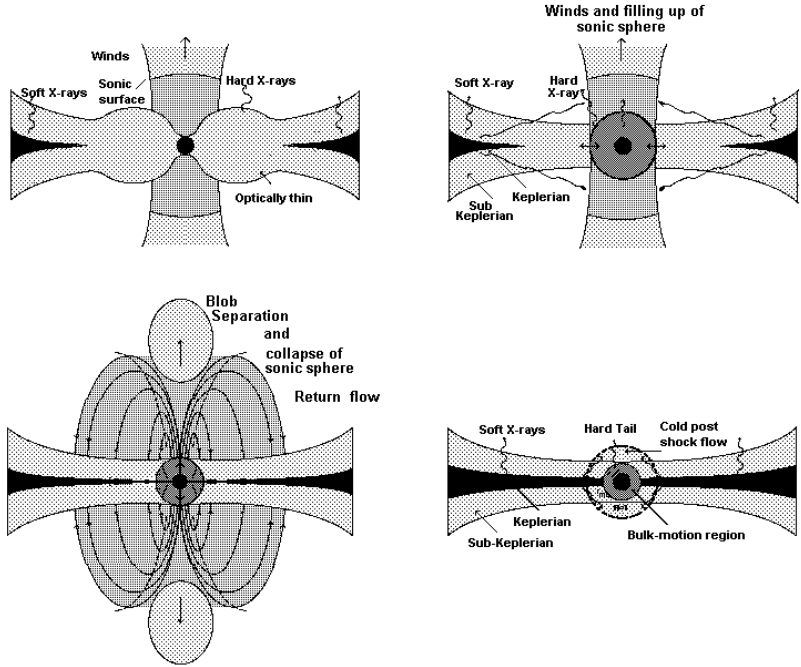}}}
\vskip 0.0cm
   \caption{Typical configurations of the two component advective flows around a black hole. See text for 
details. }
\end{center}
\end{figure}

In Fig. 4, we show schematically how the interactions among the disk, CENBOL and the jet components 
could give rise to time-dependent behaviour of the light curves (Chakrabarti et al. 2002a). Actual 
numerical simulation is difficult in presence of Comptonization, however using one type of cooling
only, we have been able to show that the Power density spectrum of the $\chi$ state could be produced as observed
(Chakrabarti, Acharyya \& Molteni, 2004; see also Chakrabarti, this volume). Details of the 
results of this non-linear analysis will be discussed elsewhere. 

\begin{figure}
   \begin{center}
{\vskip -0.5cm
   \mbox{\epsfxsize=0.9\textwidth\epsfysize=0.7\textwidth\epsfbox{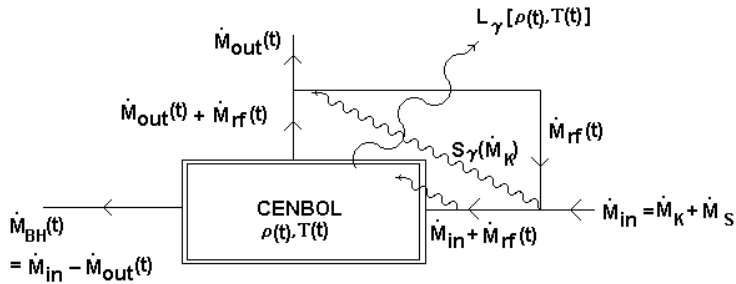}}}
\vskip -3.0cm
   \caption{Schematic diagram of the interaction of the boundary layer (CENBOL)  and the outflow with
the radiation emitted from the accretion flow. If the radiation is strong, outflow is quenched (Chakrabarti
et al. 2002b).}
\end{center}
\end{figure}

\section{Classification of light curves of GRS1915+105 and their transitions}

One of the puzzling black holes is  GRS 1915+105 which exhibits large variations in
the light curves. Originally these were classified into twelve classes
by Belloni et al. (2000) on the basis of the hardness ratio. 
Naik et al. (2002) showed that there is another independent class. 
In Chakrabarti \& Nandi (2000) it was already pointed out that these classes
are due to variations in the accretion rates. Chakrabarti et al. (2004a)  
reported some of the actual class transitions using observations with IXAE. They showed that 
these transitions are smooth and in between two `known' classes the object stays in 
an `unknown' classes. In Chakrabarti et al. (2004b), a few other class transitions 
are reported and it was further established that during a class transition, the photon 
count changes considerably and it is possible to understand this behaviour 
by assuming changes in the sub-Keplerian accretion rate. 

To begin with we present the classes as presented by Belloni et al. (2000)
in a sequence which is easier to understand using TCAF. Fig. 5 (Nandi et al. 2000)
shows the twelve classes marked by 1 ... 12 which are called  $\chi$, $\alpha$, $\nu$, $\beta$, $\lambda$, $\kappa$,
$\rho$, $\mu$, $\theta$, $\delta$, $\gamma$ and $\phi$ respectively.  Panels 3 and 6 have more than one light curve
(separated by dashed line), as they are similar but  with subtle difference. Along X-axis
is the time elapsed in seconds since the beginning of the observation. Spectral analysis of the 1st
panel suggests that it is purely in hard state. There is a prominent
QPO whose frequency may change from time to time and photon count number may also change
significantly. Final three panels (10-12) contain
light curves of those days on which spectral states are soft. There are no QPOs in these
days. Spectral fits indicate high temperature and high photon spectral index.
The ninth panel contains a light curve where two semi-soft
spectra with different photon counts are seen. Count rate varies very significantly.
In the remaining seven panels, (2-8) photons jump in between two distinct states, one with a low photon count
(off state) and the other with a high photon count (on state). 

\begin{figure}
   \begin{center}
\vskip -3.0cm
   \mbox{\epsfxsize=0.9\textwidth\epsfysize=0.8\textwidth\epsfbox{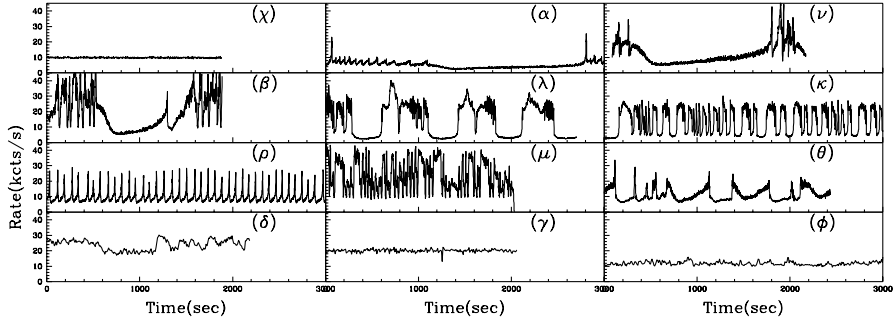}}
\vskip -4.0cm
   \caption{Classification of major light curves (Belloni et al. 2000) according to various colour-colour 
diagrams.} 
   \end{center}
\end{figure}

In Fig. 6(a-c), we present  a few examples of the class transition (Chakrabarti et al 2004b).
We present the light curves ($2-18$keV) of June 22nd, 1997 observation in the upper panel
and the mean photon index (MPI) in the lower panels. The MPI $s_\phi$ is obtained using the definition:
$s_\phi=- \frac{log(N_{6-18}/E_2) - log(N_{2-6}/E_1)}{log(E_2)-log(E_1)}$
where, $N_{2-6}$ and $N_{6-18}$ are the number of photons from the top layer of the PPC
and $E_1=4$ and $E_2=12$ are the mean energies in each channel. The Fig. 6a is in the 
so-called $k$ class. The Fig. 6b is in an unknown class and the Fig. 6c 
clearly shows the transition from the unknown class to the so-called $\rho$ class. 
The panels are separated by about three hours.

\begin{figure}
   \begin{center}
\vskip -3.0cm
   \mbox{\epsfxsize=0.8\textwidth\epsfysize=0.6\textwidth\epsfbox{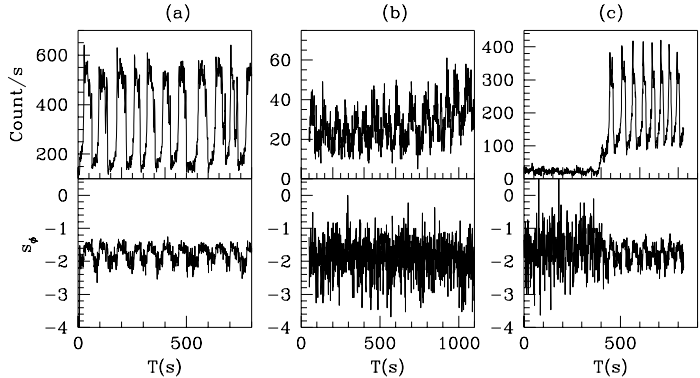}}
\vskip -2.0cm
   \caption{Light curves ($2-18$keV) as observed by IXAE (upper panel)
and the mean photon spectral index $s_\phi$ (lower panel) in the 1st, 3rd and 5th orbits
of June 22nd, 1997. GRS1915+105 was in the $\kappa$ class in (a), in an unknown class in (b)
and went to the $\rho$ class in (c) on that day. From Chakrabarti et al. (2004b).} 
   \end{center}
\end{figure}

In the lower panels, the $s_\phi$ oscillates between $\sim 2.4$ 
to $\sim 1.4$ in Fig. 6b and in Figs. 6b and 6c, the unknown class
produced very noisy photon spectral slope variation. As soon as the $\rho$
class is achieved after one `semi-$\rho$' oscillation, noise in $s_\phi$ is
reduced dramatically.

\begin{figure}
\vskip -5.0cm
\hspace {2.0cm}
   \mbox{\epsfxsize=1\textwidth\epsfysize=0.8\textwidth\epsfbox{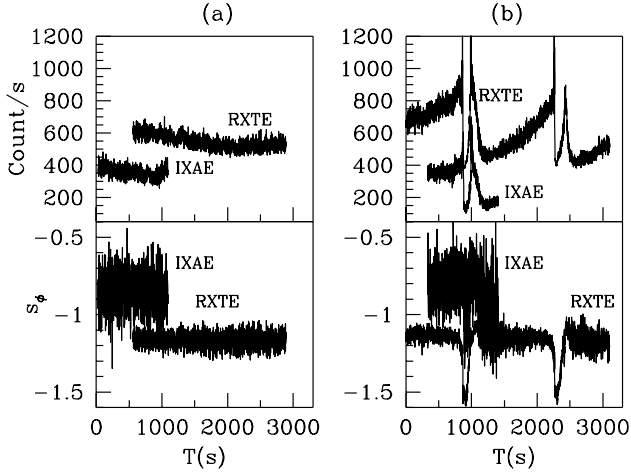}}
   \begin{center}
\vskip -2.5cm
   \caption{Class transition as seen from IXAE and
RXTE observations on the 8th of June, 1999 in two successive orbits (marked). RXTE photon counts are divided
by $50$ and shifted upward by $200/s$ for comparison. In (a), 
no significant variation in light curve or spectral index and the object was in $\chi$-like class with high counts.
In (b), the object is distinctly in the $\theta$ class. There is a gap of $44$ minutes in the 
two RXTE data presented in (a) and (b). From Chakrabarti et al. (2004b).} 
   \end{center}
\end{figure}

In Fig. 7(a-b), we show light curve and $s_\phi$ from IXAE data obtained on the 8th of June, 1999 
(Chakrabarti et al 2004a).
The two panels are from two successive orbits $\sim 80$ minutes apart. In Fig. 7a, 
the power density spectrum (PDS) is typical of that of the $\chi$ class but the count rate was very 
high compared to what is expected from such a class. A QPO at $4.7$Hz is present.
The $s_\phi$ is $0.85$  which is harder than what is observed in Fig. 6. When combined with RXTE data of that
date (Fig. 7a), one finds that for a long time ($\sim 3000$s) there was no signature of any `dip' --- 
the characteristic of the $\theta$ class.  Hence, this must be in an unknown class, more close to $\chi$
than any other. RXTE also observed this object on 7th of June, 1999 and found the
object to be in the $\chi$ class. In Fig. 7b, the light curve in the next orbit of IXAE
shows the evidence of the so-called $\theta$ class. Interestingly, the spectra gradually `hardened'
to $s_\phi \sim 0.6$ just before the `dip'. The spectra characteristically softened
in the `dip' region with $s_\phi \sim 1.4$ as the inner edge of the disk is
disappeared. This class transition is confirmed in the data of RXTE also shown
in Fig. 7b. The lower panels showed that the spectral slopes obtained for RXTE
data calculated in the similar way as $s_\phi$ was calculated before.
Here, the RXTE photons were first binned in $2-6$ keV and $6-15$keV energies before computing $s_\phi$ from
$s_\phi= - \frac{log(N_{6-15}/E_2) - log(N_{2-6}/E_1)}{log(E_2)-log(E_1)}$,
where, $E_1=4$keV, $E_2=9$keV. 

\section{Examples of possible class transitions in other objects}

Though the variability classes of other objects have not been studied 
for other objects as in GRS 1915+105, there are some evidence that the emitted
radiation changes specific characteristics in short time periods. In Fig. 8(a-b) 
we show peculiar features in RXTE light-curves (Nespoli et al. 2003) of  GX339-4 taken on MJD 52411 (May 17th, 2002)
in two successive orbits. In Fig. 8a, the light curves in the upper panel and the color (ratio of counts
in the 16.4-19.8 bands over 3.7-6.5keV bands) in the lower panel are shown. Power density spectra
indicated presence of a QPO at $\sim 6$Hz in the second orbit. Detailed study showed that there was no QPO
for first two minutes or so into of observation of the second orbit, and then the QPO appeared with
variable  frequency (Fig. 8b). The appearance of QPOs indicate the formation of an oscillating shock
(see Fig. 3, upper-right panel),
which is induced by the resonance in cooling time-scale and the infall time-scales from the post-shock flow
(Molteni, Sponholz \& Chakrabarti, 1996).
This resonance could be short lived or transient, since the accretion rate was changing during this observation.
More such observations would be essential to classify the nature of variations.

\begin{figure}
   \begin{center}
\vskip 0.5cm
   \mbox{\epsfxsize=0.4\textwidth\epsfysize=0.3\textwidth\epsfbox{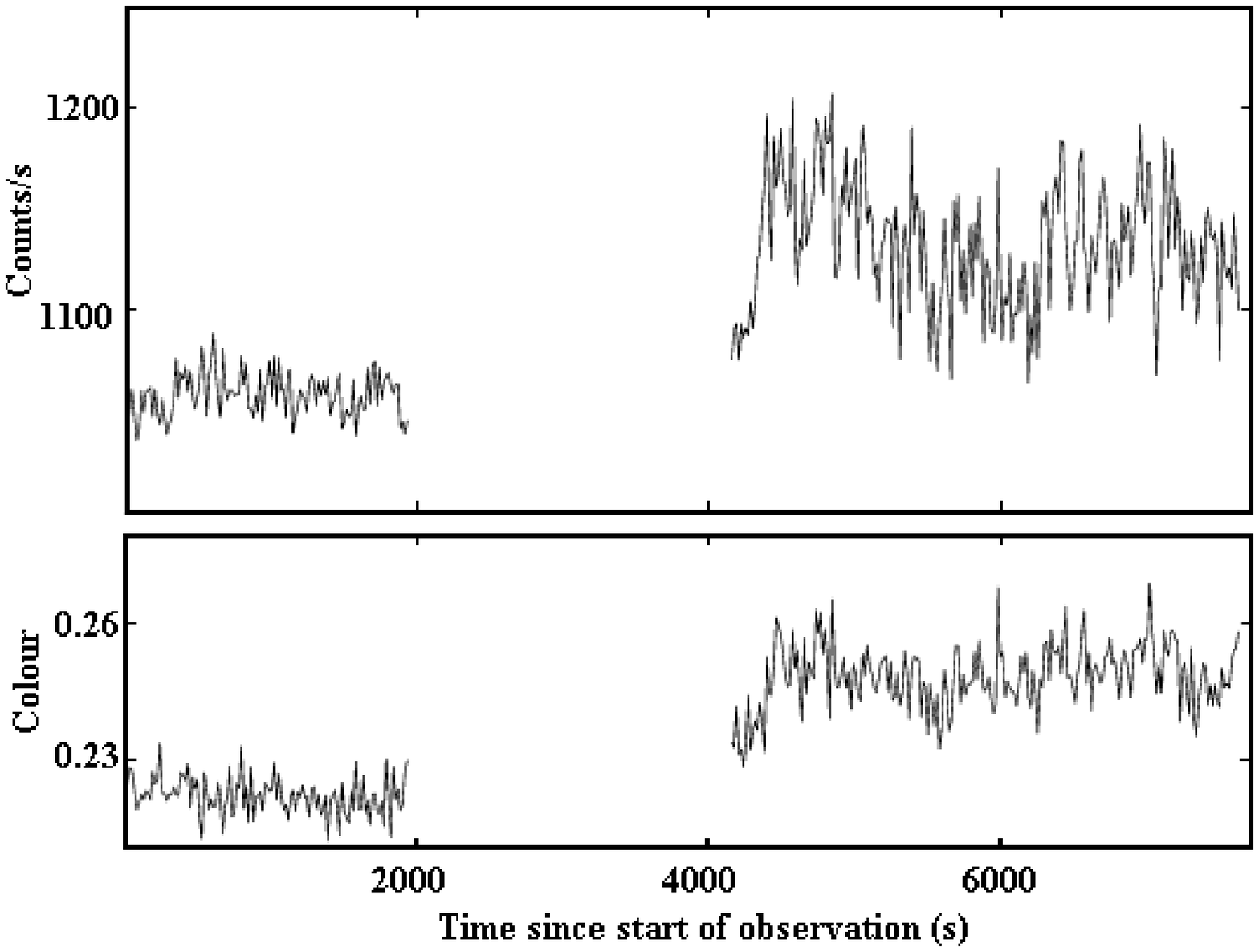}}
   \mbox{\epsfxsize=0.42\textwidth\epsfysize=0.3\textwidth\epsfbox{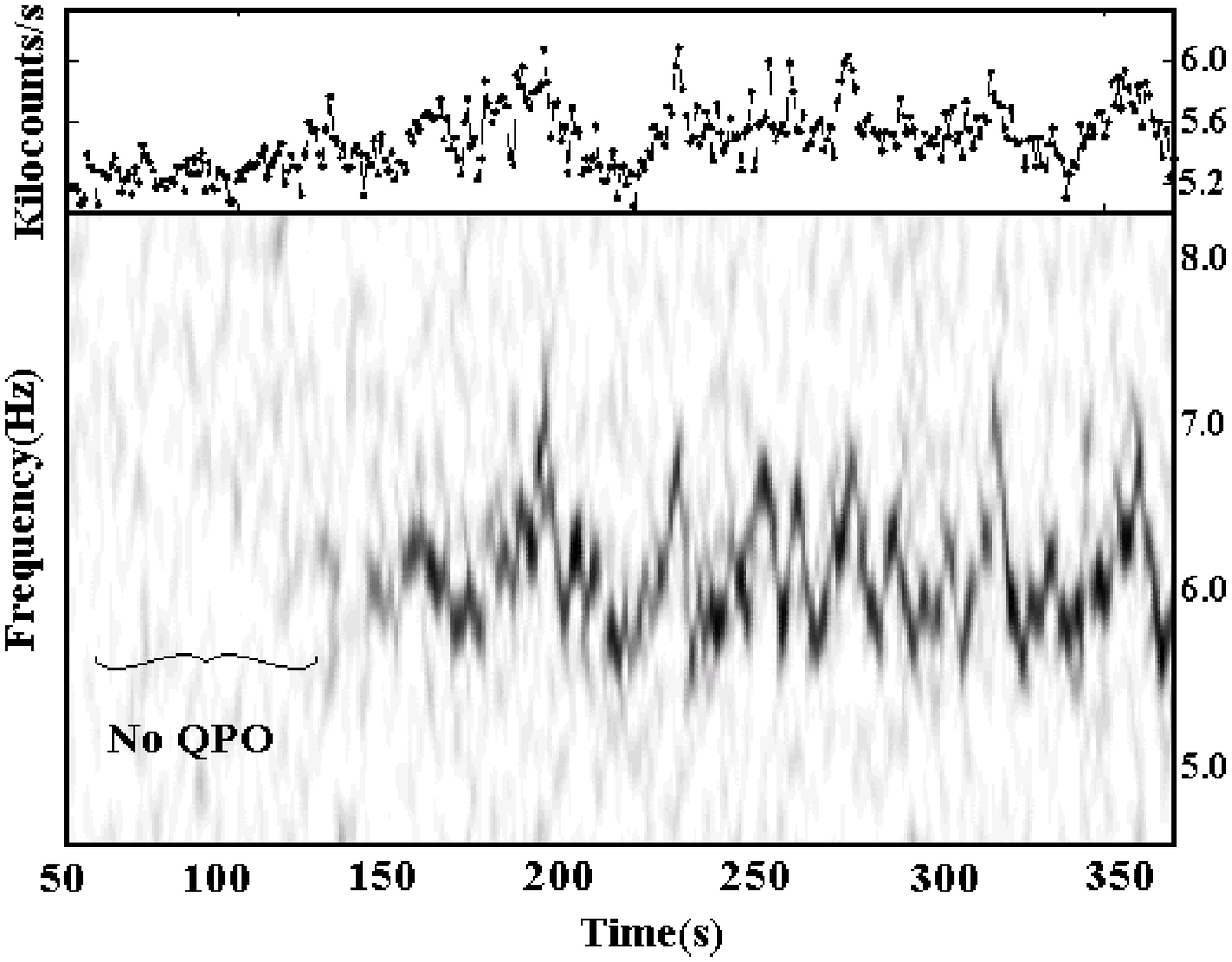}}
\vskip 0.0cm
   \caption{
Example of a transition of some kind in GX339-4 as is evident from the (a) lightcurve and  colour
variation  in two successive orbits and the (b) averaged counts and  dynamical PDS of the second orbit.
QPO appears after a couple of minutes of the begining of the second orbit indicating the change in flow
configuration from a no-shock or steady shock flow to an oscillating shock flow (Figures adapted from 
Nespoli et al., 2003).} 
   \end{center}
\end{figure}

Another example can be provided for GRO 1655-40 
in which the state transition has been observed (Sobczak et al. 1999, Remillard et al. 1999). 
Fig. 9(a-b) shows the spectra at four times with the following PIDs:
(i) 20402-02-02-00 (Mar. 05, 1997) to represent high/soft state (lower-right panel of Fig. 3),
(ii) 20402-02-25-00 (Aug. 14, 1997) to represent an intermediate hard-state
which is lossing matter to the sonic sphere (early stage of lower-left panel of Fig. 3),
(iii) 20402-02-24-00 (Aug. 03, 1997) to represent an intermediate soft state just before 
the state transition (late stage of lower-left panel of Fig. 3)
and finally (iv) 20402-02-26-00 (Aug. 18th, 1997) to represent a true hard state (upper right panel of Fig. 3). 
We fitted the spectra keeping the hydrogen column density fixed at $0.89 \times 10^{22}$ atoms 
per $cm^{-2}$ (Zhang et al. 1997). 
According to the theoretical understanding (Chakrabarti, 1999; 2002b), one should expect case 
(iii) to have harder spectra compared to case (i) and case (ii) to have a softer 
spectra compared to case (iv). That is precisely seen in Figs. 9a and 9b respectively. 
There is no evidence for QPO in both the soft states (Fig. 9a) while a strong QPO is observed 
in case (ii) at 1.4Hz and 6.4Hz while a weak QPO is observed in case (iv) at 0.2Hz and 0.8Hz (Fig. 9b).
If oscillation of shocks correspond to QPOs (Molteni, Sponholz \& Chakrabarti, 1996;
Chakrabarti \& Manickam, 2000) {\it and} if shocks also produce outflows
(Chakrabarti, 1999) then it is clear that a strong shock (and therefore outflow)
is present in case (ii) while there is a weak shock very far out (since
$\nu_{QPO} \propto R_s^{3/2}$, where $R_s$ is the shock location) in case (iv).

\begin{figure}
\vskip -4.0cm
\hskip 1.5cm
\mbox{\epsfxsize=1.0\textwidth\epsfysize=0.8\textwidth\epsfbox{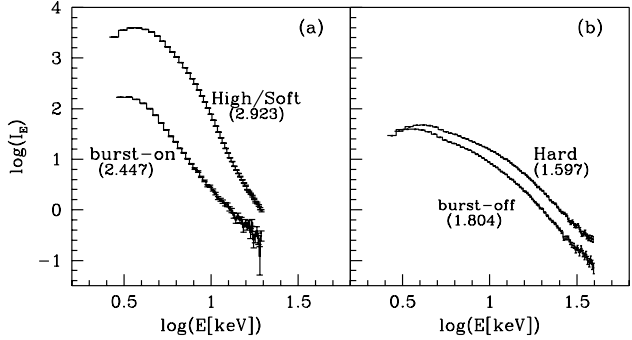}}
{   \begin{center}
\vskip -4.0cm
   \caption{Comparison of spectra of (a) high and burst-on and (b) low and burst-off phases of 
GRO 1655-40 clearly showing hardening of the soft state and softening of the hard state.}
   \end{center}}
\end{figure}

The softening of the hard state and the hardening of the soft state spectra have been 
found to be related to the presence of outflow and return flow respectively (Chakrabarti
et al. 2002b). These phenomena occur because when matter goes out of the CENBOL
the number of electrons go down for the same source of soft photons and the spectrum is 
softened. Conversely, when matter returns back, the spectrum is hardened. as a result,
the pivotal point between the burst-on and burst-of states (where the wind activity is
prominent) shifts outward  by a few keV as compared to the pivotal point between the 
hard and soft states (when the wind activity is relatively lower). This phenomenon is 
routinely observed in GRS 1915+105. Figs. 10(a-c) show the comparison of spectra 
in burst-on and burst off states of (a) June 18, 1997,  (b) July 10, 1997 and (c) July 12, 1997
respectively. The spectra in the high and  the low states for the same object on August 19, 1997
and March 26, 1997 respectively are plotted to establish the shifting of the pivotal point
in presence of winds.

\begin{figure}
   \begin{center}
\vskip -1.0cm
   \mbox{\epsfxsize=1.0\textwidth\epsfysize=0.8\textwidth\epsfbox{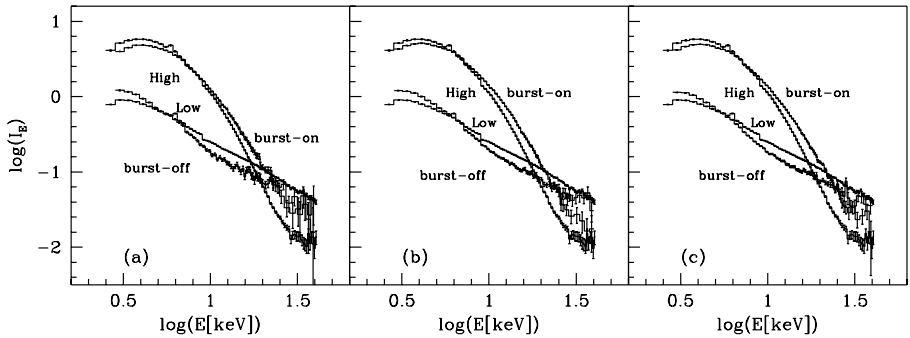}}
\vskip -8.0cm
   \caption{Comparison of the spectra of GRS 1915+105 on three different days with generic high (typically $\phi$)
and low (typically $\phi$) state spectra of the same source to indicate that variability class transition to 
burst-off/burst-on (typically, $\lambda$ or $\kappa$) is also 
accompanied by shifting of the pivotal energy. From Chakrabarti et al (2002b). } 
   \end{center}
\end{figure}

\section{Concluding Remarks}

Variability class transitions are common in GRS 1915+105 and probably have been seen in GX 339-4 and GRO 1655-40.
For a given black hole, there are only two parameters, namely, the accretion rate of the Keplerian component 
and that of the sub-Keplerian halo which can give rise to these transitions. Alternatively, if one assumed that
both the matter is supplied at the outer edge of the disk, then this accretion rate together with the
viscosity parameter ($\alpha(r,z)$) should be enough. In the TCAF solution, the separation of the Keplerian
from sub-Keplerian can be achieved by appropriate viscosity parameters, and hence independent 
variation of the two rates produces the same result.

We presented a few examples of the class transitions in GRS 1915+105 especially using the IXAE experiment 
abroad IRS P3. The count rate was found to change before and during class transition
and the two classes before and after the transition were not found to have `normal' counts characteristics of 
those classes. The transition is also through an unknown class. It takes about 2-3 hours for a transition 
which may typically be the free-fall time-scale of the low-angular momentum (sub-Keplerian) flow. Thus,
it is possible that the major cause of transition is the slow change in the accretion rates.

The transitions discussed in this review are related to the galactic black holes only. For extra-galactic black holes,
the same phenomenon is expected, but on a quite longer time-scale (as each time interval scales with the 
mass of the central hole) and it is not clear if this has been observed yet.

\acknowledgements 
This work was partly supported by the DST project `Emitted Spectra from Two-Component 
Accretion Disks Around Black Holes'.

\label{lastpage}


\begin{thebibliography}{99}

\bibitem[]{}
Belloni, T. et al., 2000 A \& A 355 271

\bibitem[]{}
Chakrabarti, S.K. 1990, Theory of Transonic Astrophysical Flows (World Scientific: Singapore)

\bibitem[]{}
Chakrabarti, S.K. 1993, Numerical Simulations in Astrophysics, J. Franco et al.  (Eds.),
Mexico city, Mexico (Cambridge University Press:UK)

\bibitem[]{}
Chakrabarti, S.K. 1994, Proceedings of 17th Texas Symposium In  H. B\"ohringer et al. (Eds.),
Munich, Germany, New York Academy of Sciences, New York

\bibitem[]{}
Chakrabarti, S.K. 1996, Physics Reports, v.266, No 5 \& 6, 229

\bibitem[]{}
Chakrabarti, S.K. 1998, Ind. J. Phys., 72(B), 565

\bibitem[]{}
Chakrabarti, S.K. 1999, Astron. \& Astrophys., 351, 185

\bibitem[]{}
Chakrabarti, S.K. 2001, Astrophys. and Space Science, 276, 191

\bibitem[]{}
Chakrabarti, S.K., Acharyya, K. \& Molteni, D. 2004a, A \& A 421, 1

\bibitem[]{}
Chakrabarti, S.K. \& Manickam, S.G. 2000, ApJ 531, L41

\bibitem[]{}
Chakrabarti, S. K., Manickam, S. G., Nandi, A. and Rao, A. R. 2002, 
in the Proceedings of the IXth Marcel Grossman Meeting, Eds. V. G. Gurzadyan, R. T. Jantzen, 
R. Ruffini, (World Scientific Co.: Singapore) 2279

\bibitem[]{} 
Chakrabarti, S.K. \&  Molteni, D. 1995, M.N.R.A.S., 272, 80

\bibitem[]{}
Chakrabarti, S.K. \& Nandi A. 2000, Ind. J. Phys., 75(B), 1

\bibitem[]{}
Chakrabarti, S.K., Nandi, A.,  Chatterjee, A. K.,  Choudhury, A., Chatterjee, U., 2004b, A\&A (to appear)

\bibitem[]{}
Chakrabarti, S.K., Nandi, A., Manickam, S.G., Mandal, S. \& Rao, A.R., 2002b, ApJ, 579, L21

\bibitem[]{}
Chakrabarti, S.K. \& Titarchuk, L.G. 1995, ApJ, 455, 623

\bibitem[]{}
Molteni, D., Sponholz, H. \& Chakrabarti, S.K. 1996, ApJ 457, 805

\bibitem[]{}
Naik, S., Rao, A.R. \& Chakrabarti, S.K. 2002a, J. Astron. Astrophys, 23, 213 

\bibitem[]{}
Nespoli, E., et al., 2003,  A\&A 412, 235 

\bibitem[]{}
Remillard, R.A., Morgan, E.H., McClintock, J.E., Bailyn, C.D. and Orosz, J.A.,
1999, ApJ, 522, 397

\bibitem[]{}
Shapiro, S.L., Lightman, A.P. \& Eardley, D.M. 1976, ApJ, 204, 187 

\bibitem[]{} 
Smith, D.M., Heindl, W.A. \& Swank, J.H. 2002, ApJ, 569, 362

\bibitem[]{}
Sobczak, G.J., McClintock, J.E., Remillard, R.A., Bailyn, C.D. and Orosz, J.A.,
1999, ApJ, 520, 776

\bibitem[]{}
Zhang, S.N., et al. 1997, ApJ, 479, 381

\end{thebibliography}
\end{document}